\documentclass[11pt, a4paper]{article}
\usepackage{array}
\usepackage{graphicx}
\usepackage{amsfonts}
\usepackage{amssymb}
\usepackage{amsthm}
\usepackage{amsmath}
\usepackage{multirow}
\usepackage{color}
\usepackage{cite}
\usepackage{float}
\usepackage{geometry}
\usepackage{authblk}
\usepackage{hyperref}
\usepackage{cleveref}
\usepackage{amsmath}   
\usepackage{amssymb}   
\usepackage{siunitx}   

\newcommand{\nn}{{\nonumber}}

\newcommand{\beq}{\begin{equation}}
	\newcommand{\eeq}{\end{equation}}
\newcommand{\bea}{\begin{eqnarray}}
	\newcommand{\eea}{\end{eqnarray}}

\newcommand{\gsim}{\lower.7ex\hbox{$\;\stackrel{\textstyle>}{\sim}\;$}}
\newcommand{\lsim}{\lower.7ex\hbox{$\;\stackrel{\textstyle<}{\sim}\;$}}
\newcommand{\be}{\begin{equation}}
	\newcommand{\ee}{\end{equation}}
\newcommand{\ba}{\begin{eqnarray}}
	\newcommand{\ea}{\end{eqnarray}}

\usepackage{caption}

\newcommand{\hnk}{\hat{n}_k}
\newcommand{\hnz}{\hat{n}_0}

\title{ 5D Rotating Black Holes as dark matter in Dark Dimension Scenario: Hawking Radiation versus the Memory Burden Effect}
\author{George K. Leontaris\thanks{Email: \texttt{leonta@uoi.gr}}\;}
\author{\,George Prampromis\thanks{Email: \texttt{g.prampromis@uoi.gr}}}

\affil{Physics Department, University of Ioannina, 45110, Ioannina, Greece}

\begin{document}
	
	\maketitle

	\begin{abstract}
		    This work explores the possibility that five-dimensional primordial rotating black holes could account for all, or a significant portion, of the dark matter in our universe. Our analysis is performed within the context of the ``dark dimension'' scenario, a theoretical consequence of the Swampland Program\cite{Vafa:2005ui} that predicts a single micron-scale extra dimension to explain the observed value of dark energy. 		
		We demonstrate that within this scenario, the mass loss of a primordial rotating black hole sensitive to the fifth dimension is significantly slower than that of its four-dimensional counterpart. Consequently, primordial black holes with an initial mass of $M\gtrsim  10^{10}$g can survive to the present day and potentially constitute the dominant form of dark matter.		
	Finally, we investigate the memory burden effect and find that it dramatically prolongs the lifetime of five-dimensional rotating primordial black holes, making them compelling candidates for all the dark matter in the universe.

	\end{abstract}

	\newpage

	\section{Introduction}

	The string landscape comprises a vast collection of potentially stable or metastable four dimensional (4D) vacua, arising from compactifying the extra dimensions on Calabi-Yau manifolds. 
	The derivation of 4D physics from string theory is achieved through various compactification schemes, notably those incorporating background fluxes and D-brane configurations. Critically, each stabilized vacuum corresponds to a different low-energy effective field theory, fundamentally defined by its unique set of parameters, including coupling constants, gauge groups, and matter spectra. This mechanism is the very origin of the vast string landscape.
	Thus, given the enormity of this landscape, an additional principle is necessary to identify vacua that correspond to our observable universe. The Swampland Program addresses this issue by proposing a set of conjectured constraints on EFTs that can be consistently coupled to quantum gravity\cite{Vafa:2005ui,Ooguri:2006in}. Its objective is to delineate the landscape of consistent quantum gravity theories from the ``swampland'' of seemingly effective theories that are incompatible with a UV-complete formulation, such as string theory~\cite{Palti:2019pca,vanBeest:2021lhn}. 

	A challenging  issue in EFTs is the naturalness of the  hierarchy of energy scales. Regarding  cosmology  in particular, a key point at issue is the smallness of dark energy scale, which in the gravitational field equations is expressed by the cosmological constant $\Lambda$. Cosmological data show that its value today is
	\be 
	\Lambda = 10^{-122}  M_{P}^4~.\label{LambdaValue}
	\ee 
	Recent progress~(see e.g. review~\cite{Agmon:2022thq}) provided the 
	tools to make a leap forward in understanding its origin. 
	From the perspective of swampland conjectures -particularly the Distance Conjecture and the Weak Gravity Conjecture~\cite{Ooguri:2018wrx,Montero:2022prj}- the  scale $\Lambda$   has significant implications. 
The Distance Conjecture~\cite{Ooguri:2018wrx}, in particular, states that in any effective field theory consistently coupled to gravity, a large excursion $\Delta\phi\gg 1$ (in Planck units) in field space for a modulus $\phi$ implies the emergence of an infinite tower of states with masses exponentially suppressed as
\begin{equation}
	m \sim m_0 e^{-\alpha \Delta\phi}, \quad \alpha \sim \mathcal{O}(1)~.
\end{equation}

Remarkably, this can be adapted to an Anti–de Sitter (AdS) background. The AdS Distance Conjecture~\cite{Lust:2019zwm} postulates that as the cosmological constant $\Lambda <0$  approaches zero, a tower of states becomes light, with a mass scale related to $|\Lambda|$. In particular, for the Kaluza–Klein (KK) scale $m_{KK}$ one finds that it is connected to the cosmological constant as follows

\be 
m_{KK}\sim |\Lambda|^a,\, a=\frac 14\label{KKLambda}
\ee 
Given the observed tiny value of  $|\Lambda|$ in Planck units, this relation predicts a KK-scale of the order  $m_{KK}\sim \mathcal{O}$(eV), corresponding to an extra dimension of $\mu$m size.

Further arguments suggest that a similar scaling holds in de Sitter (dS) space ($\Lambda >0$), with a tower of light states emerging as $\Lambda\to 0$. Taken together, these conjectures motivate the Dark Dimension scenario, which correlates the dark energy scale $\Lambda$ with the size $R_C$ of a compact extra dimension. Combining theoretical reasoning with observational constraints leads to the relation
\be 
R_C \sim \lambda \Lambda^{-\frac 14}~,
\ee 
where the numerical prefactor lies in the range $\lambda\in[10^{-1}-10^{-4}]$.

Various theoretical and experimental constraints limit the parameters and mass scales of the scenario. An important consideration arises from the massive spin-2 bosons contained within the KK tower. Higuchi’s analysis~\cite{Higuchi:1986py}  of free massive spin-1 and spin-2 fields in the expanding part of de Sitter spacetime imposes a strict upper limit on the exponent $a$ in the scaling relation $m\sim \Lambda^a$ restricting it to the range $\frac 12 \ge a \ge \frac 14$.
However, fifth-force experiments further constrain the geometry of the compactification space as they allow only a single extra dimension in the micron-scale range. This effectively saturates the lower bound of $a$, fixing it to $a=1/4$~\cite{Montero:2022prj,Anchordoqui:2023laz}. 
%
%
Furthermore, it is known that in the presence of  extra compact dimensions there are modifications in Newton's inverse square law~\cite{Floratos:1999bv,Kehagias:1999my}.   Experimentally this implies that there can be only one micron-size extra dimension, since Newton's law has been  verified~\cite{Adelberger:2003zx,Lee:2020zjt} down to 30 $\mu$m.  These constraints~\footnote{Constraints on the number of  micron-size extra  dimensions also arise from neutron star heating~\cite{Hannestad:2003yd,Lehnert:2025izp}. For extra dimensions of  ${\mu}m$ size, only   $n=1$ is  compatible (see however ~\cite{Anchordoqui:2025nmb} for the case $n=2$).}. therefore imply a lower bound on the characteristic mass scale of the emerging KK tower~\cite{Montero:2022prj}
\be 
m\sim m_{KK} \gtrsim 6.6 {\rm meV} 
\ee

 The Standard Model of the electroweak and strong interactions can reside on a D-brane  transverse to the fifth dimension~\cite{Antoniadis:1997zg} while physics can still be  described by a higher-dimensional Quantum Field Theory (QFT) up to the species scale $\Lambda_{\text{sp}}$~\cite{Dvali:2007hz} which is of the order of the higher-dimensional Planck scale $  M_{P}^2\sim M_{*}^{(n+2)} R_C^n$,  where $R_C\sim\frac{1}{m}$ is the size of extra dimension which is of the order of $ 1 \mu$m.  Beyond this scale the physics becomes strongly coupled to gravity and the effective theory breaks down. Thus, for the  four-dimensional macroscopic space and $n$ decompactified dimensions
	the scale up to the point where the effective description is meaningful  is given by 
	
	\be 
	\Lambda_{sp} = M_{*}\approx \left(m^n\,M_{P}^2\right)^{\frac{1}{n+2}}
	\label{Pmass}
	\ee

	For one extra dimension ($n=1$),  and a KK-tower of mass $m_{KK}\sim \frac{1}{R_C}$, the species scale is 
	\[ \Lambda_{sp}\sim  (m_{KK} M_P^2)^{1/3}     \]

 Then, the five-dimensional species scale is of the order
	\be 
	\Lambda_{sp} 
	= M_*\sim  \left(m\,M_{P}^2\right)^{\frac{1}{3}}
	\sim  10^{10}{\rm GeV}	\label{Pmass1}
	\ee

	The Dark Dimension scenario has profound implications for the physics of Primordial Black Holes (PBHs), particularly concerning their potential role as a dark matter  component. Indeed, PBHs, formed from the gravitational collapse of primordial overdensities \cite{ZeldovichNovikov1967,Hawking1971,CarrHawking1974,Chapline:1975ojl,Meszaros:1975ef}, have been extensively investigated for decades as a candidate for dark matter (DM).

	In the standard four-dimensional framework, PBHs with masses $M\lesssim 5\times 10^{14}$ g would have completely evaporated by the present day due to Hawking radiation and thus cannot be DM  candidates. For higher masses, stringent bounds, primarily from the extragalactic gamma-ray background generated by their evaporation, severely constrain their abundance~\cite{Ballesteros:2019exr}. These observations effectively rule out the possibility that PBHs in the range $\sim 10^{15} {\rm g} \lesssim M_{PBH} \lesssim 10^{17}$g could  constitute all dark matter. While the constraints are somewhat weaker for more massive PBHs, other limits from microlensing~\cite{Mediavilla:2017bok} make it highly improbable for PBHs in any sub-range to account for 100\% of the dark matter.
	
	Recent  works~\cite{Anchordoqui:2022txe,Anchordoqui:2024jkn,Anchordoqui:2024akj,Anchordoqui:2024tdj} explored whether PBHs within the Dark Dimension scenario could account for all DM, focusing on non-rotating  black holes and other related issues~\cite{Anchordoqui:2025opy},   (see also \cite{deFreitasPacheco:2023hpb} for Quasiextremal primordial black holes and the review~\cite{Carr:2020xqk}). The analysis of~\cite{Anchordoqui:2022txe,Anchordoqui:2024jkn,Anchordoqui:2024akj,Anchordoqui:2024tdj} in particular indicates that Hawking radiation is suppressed in higher dimensions, allowing lighter PBHs to survive until the present epoch and thereby contribute to the DM density.
	Furthermore, suppressed Hawking emission implies fewer emitted particles, and therefore, constraints from the CMB, galactic bulge positrons, and isotropic photon backgrounds~\cite{Poulter:2019ooo,Iguaz:2021irx,DeRocco:2019fjq} can be relaxed.

	In this work we extend the analysis of PBH evaporation to the case of rotating  black holes~\cite{Kerr:1963ud} in a five-dimensional (5D) scenario, where the observable universe is described as a 4D brane living in a 5D higher-dimensional space. Rotating PBHs undergo two concurrent processes: Hawking radiation reduces their mass,  while evaporation cause them to eliminate  its angular momentum, ultimately transitioning to a non-rotating, Schwarzschild-like phase.	Assuming a single decompactified extra dimension on the scale of a few micrometers, we compute the complete evaporation dynamics of a 5D rotating black hole. Our analysis begins with the spin-down phase, where we calculate the mass fraction remaining and the time required for the black hole to lose its angular momentum. We then deal the subsequent evaporation and the lifespan of this resulting Schwarzschild-phase black hole.
	
	By comparing the total evaporation lifetime against the age of the universe, we assess the viability of  PBHs as DM candidates. We find that the spin-down timescale is of the order of $\mathcal{O}(10^9)$ years, comparable to the evaporation timescale~\cite{Anchordoqui:2024akj} of Schwarzschild PBHs. The combined lifetime of both phases results in additional enhancement, which corresponds to a lower mass bound of approximately $10^{10}$ grams for a PBH to survive as a DM candidate.
	
	We further incorporate the recent ``memory burden'' effect proposed by Dvali~\cite{Dvali:2018xpy}. According to this scenario, as  black hole evaporates, it retains quantum information from absorbed matter, leading to a growing informational load that modifies its late stage evolution. We demonstrate that such quantum effects dominate roughly after the half-decay time, rendering semiclassical evaporation models inadequate. Our  estimates show that the memory burden suppresses the Hawking radiation rate, significantly prolonging PBH lifetimes and allowing lighter black holes to persist as DM.

	This work is structured as follows. We begin in section 2 by establishing the  framework for higher-dimensional black holes and summarizing experimental mass constraints. section 3 details the relevant rotating black hole solutions and presents our calculations of their greybody factors. 
	In section 4	we solve the coupled evolution for the black hole's mass $M$ and angular momentum $J$, obtaining an analytical relation $M(J)$. This reveals that a black hole loses roughly $\sim 50\%$ to 60\% of its mass by the time its spin vanishes. Furthermore,	 we compute the mass loss for the rotating black hole and  show that this is 	a process occurring over $\sim 10^9$ years. Finally, in section 5, we demonstrate that incorporating the memory burden effect significantly extends the black hole lifetime. This prolonged decay timescale allows rotating higher-dimensional black holes to potentially constitute all the dark matter observed in the universe.

	\section{A few basic facts about  Higher-Dimensional Black Holes} 
	It has long been argued~\cite{ZeldovichNovikov1967,Hawking1971, CarrHawking1974}  that the primordial black holes (PBHs) were 
formed when large-amplitude perturbations from the primordial curvature spectrum re-entered the Hubble horizon.  These perturbations are thought to originate from an enhancement of small-scale power in the primordial curvature spectrum generated during inflation. This spectrum, denoted $P_{\cal R}(k)$, quantifies the variance of these perturbations as a function of wavenumber $k$. On cosmological scales, such as those probed by the cosmic microwave background (CMB), it is conventionally parametrized as:

\[   P_{\cal R} (k) = A_s \left(\frac{k}{k_*}\right)^{n_s-1 }\]
Here, $k$ is the wavenumber (spatial frequency),  $k_*$ the pivot scale (reference wavenumber $=0.05 {\rm Mpc}^{-1}$), $A_s$ the amplitude of scalar perturbations, ($A_s\approx 2.1\times  10^{-9}$),  $n_s$ the spectral index (tilt), and $\alpha_s$ the running of spectral index $=dn_s/d\ln k$.
Cosmological data tightly constrain $A_s$ and $n_s$  at these scales, limiting primordial fluctuations. However, at small scales ($k\gg k_*$) constraints are weaker due to the lack of direct observations. As stressed above, amplitudes$~\sim 10^{-2}$ are observationally permitted and can induce PBH formation	
via gravitational collapse~\cite{ZeldovichNovikov1967, CarrHawking1974}.

Once PBHs form, they typically possess non-zero angular momentum $J\ne 0$, however, over extremely long timescales, $J$ can be significantly reduced through  Hawking radiation or accretion, potentially leading to a phase resembling a Schwarzschild black hole. 

The masses of the black holes vary in a wide range. 
A	connection between the PBH mass and the horizon mass at formation gives an estimate of the mass~\cite{Carr:2020xqk}
\be 
M_{BH}\sim \frac{c^3 t}{G}\approx 10^{38} \left(\frac{t}{s}\right) g=  10^{15} \frac{t}{10^{-23} s}\,g~,
\ee 
where time is in seconds and $s,g$ denote seconds and grams.

PBH masses span a wide range. Through Hawking radiation, they emit particles and lose most of their mass. To estimate PBH lifetimes,  the relevant greybody factors  must be computed for each particle species, accounting for how the gravitational potential scatters or absorbs radiation. 
However, due to various constraints derived from observations of galactic as well as extra-galactic $\gamma$-rays,  PBHs can contribute to DM only in the following range
\[  10^{15}g\lesssim M_{BH}\lesssim  10^{21}g~.\]

\begin{figure}[h]
		\centering
		\includegraphics[width=0.75\textwidth]{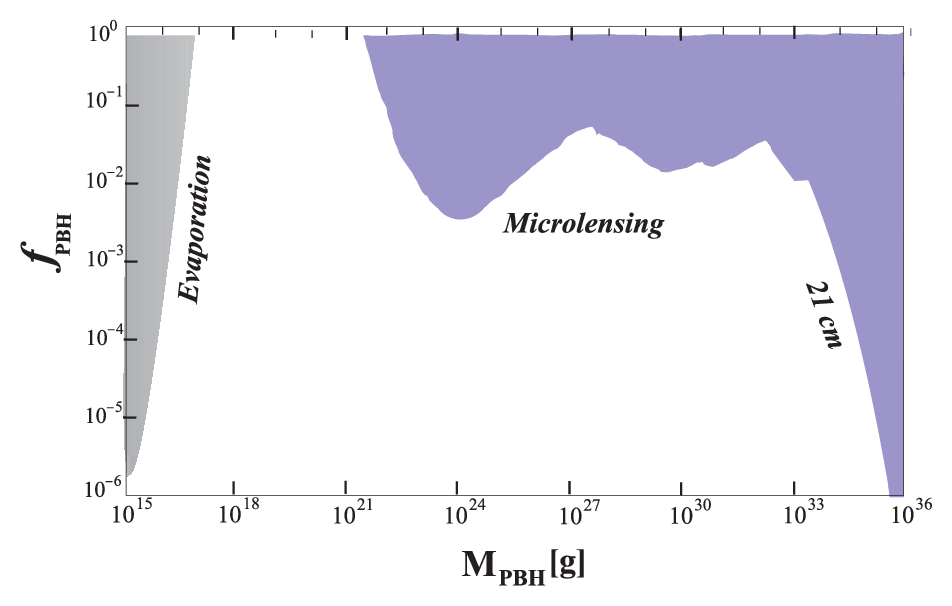}\\
		\caption{ Constraints on $ f_{PBH}$ in 4D as a function of the PBH mass $ M_{BH}$}
		\label{PBHlimits}
	\end{figure}

The black hole mass distribution can be expressed in terms 
of the BH mass $M_{BH}$ and the rate of change  of their
density number $n_{BH}$  with respect to  $M_{BH}$ in the range $M_{BH},M_{BH}+d M_{BH}$
\be 
\psi=\frac{M_{BH}}{\rho_{CDM}} \frac{d n_{BH}}{d M_{BH}}~,
\ee 
where 	$\rho_{CDM}$ stands for the cold dark matter energy density.
The total fraction $f$ of BH dark matter  is determined by the integral  
\be 
f=\int \psi d M_{BH}=\frac{\rho_{PBH}}{\rho_{CDM}}\le 1,\;{\rm where}\,,
\rho_{PBH}=\int M_{BH}dn_{BH}~.
\ee
Clearly, if $f=1$ then all dark matter is made only of PBHs. The upper limits on the allowed fraction $f$   of PHBs composed from available data are shown in figure~\ref{PBHlimits} (for retailed bounds see~\cite{Villanueva-Domingo:2021spv}).

	\subsection{ Higher Dimensional BHs}
In the previous section we saw why PBHs are not sufficient to explain the total dark matter content in the standard four-dimensional framework. In this section we revisit this conclusion within the dark dimension scenario, assuming that  black holes are situated on a D3-brane alongside the particles of the Standard Model.

	The spacetime geometry of a higher-dimensional black hole  depends critically on the distance scale relative to the compactification radius, $ R_C $, of the extra dimensions.
	At distances much smaller than the compactification scale,  $ r \ll R_C $, 
	the gravitational field is free to propagate in all $4 + n $ dimensions. Consequently, the BH's geometry is approximately that of a higher-dimensional Schwarzschild-Tangherlini~\cite{Tangherlini:1963bw} black hole, with a metric  given by:
	\begin{equation}
		ds^2 = -h(r) dt^2 + \frac{dr^2}{h(r)} + r^2 d\Omega_{n+2}^2~.
	\end{equation}
	The function $h(r) = 1 - \left( \frac{r_H}{r} \right)^{n+1} $ encodes the gravitational potential, and $ r_H $ is the higher-dimensional horizon radius. The term $ d\Omega_{n+2}^2 $ represents the metric on a unit $ (n+2) $-sphere, describing the geometry of the angular dimensions and it is given by,
	\[
	d\Omega_{n+2}^2 = d\theta_{n+1}^2 + \cdots + \sin^2\theta_{n+1} \left( d\theta_n^2 + \cdots + \sin^2\theta_2(d\theta_1^2 + \sin^2\theta_1 d\phi^2) \right)~.
	\]

	At distances much larger than $ R_C$, the extra dimensions are effectively ``undetected'' and the gravitational field is forced to propagate only along the four-dimensional spacetime. The metric  therefore acquires the standard four-dimensional Schwarzschild form:
	\begin{equation}
		ds^2 = -\left(1-\frac{2 G M_{\text{BH}}}{r}\right)dt^2 + \frac{dr^2}{1-\frac{2 G M_{\text{BH}}}{r}} + r^2(d\theta^2 + \sin^2\theta d\phi^2)~.
	\end{equation}

	Before we proceed to the implementation of the scenario,  we recall a few relevant results of the higher dimensional Schwarzchild solution~\cite{Tangherlini:1963bw}, following the analysis~\cite{Anchordoqui:2024jkn}. 
	We work in the limit where all distance processes shorter that the Planck length can be neglected in BH production.
	More precisely, we are interested in BHs with Schwarzchild radius $r_S$ in the range
	\be 
	\ell_s \ll r_S \ll R_C~,\label{sizebound}
	\ee
	where $\ell_s$ is the string scale and $R_C$ as previously the compactification radius of the extra dimension. In this framework, 
	we consider black holes radiating mostly  on the 4D brane, 
	hence, we  need to compute the probability distribution of the outgoing particles on the brane.

	 The thermal spectrum in $4+n$ dimensions is
	\be 
	T_{BH}= \frac{n+1}{4 \pi r_S}~,\label{TempBH}
	\ee 
	where the radius $r_S$ is
	\cite{Kanti:2002nr}
	\be  
	r_S= \frac{1}{M_{*}}\left(\frac{M_{BH}}{M_{*}}\right)^{\frac{1}{n+1}}
	\left(\frac{8\Gamma(\frac{n+3}2)}{(n+2)\pi^{\frac{n+1}2}}\right)^{\frac{1}{n+1}}
	=
	\frac{1}{\sqrt\pi \,M_{*}}\left(\frac{M_{BH}}{M_{*}}\right)^{\frac{1}{n+1}}
	\left(\frac{8\Gamma(\frac{n+3}2)}{n+2}\right)^{\frac{1}{n+1}}
	\label{rsradius}
	\ee 
	
	Also, the entropy in the case of $n$ extra dimensions is given by
	\be 
	S=
	\frac{4\pi M_{BH}r_S}{n+2}\;\equiv\;\frac{n+1}{n+2}\frac{M_{BH}}{T_{BH}}
	\label{EntropyN} 
	\ee

	Having established the essential features of higher-dimensional black hole physics, we now address the central focus of this paper, and in particular  their cosmological implications and the possibility of being a substantial dark matter  component.
	We start the next section  by providing the basic results on Hawking radiation and lifetime of the 5-dimensional Schwarzchild BH.

	\subsection{Lifetime of the Schwarzchild BH }
	
We first summarize the results for the Schwarzschild case~\cite{Anchordoqui:2022txe} before presenting our computations for the rotating black hole.
In this work, we consider a black hole localized on the brane. Given that the standard model degrees of freedom are confined to the brane, with only gravitons coupling to the bulk spacetime, the primary emission channel for Hawking radiation is anticipated to be on the brane. We therefore concentrate our subsequent analysis and calculations on this dominant brane emission.


	The following expression governs the time dependence of the black hole mass loss rate due to Hawking radiation:
	\begin{equation}
		-\frac{dM}{dt} = \sum_{s, l, m} \int g_s \frac{d\omega}{2\pi} \frac{\Gamma_{s,l,m}(\omega, M) , \omega}{\exp\left(\omega/T_H\right) - (-1)^{2s}} ~,
	\end{equation}
	where $g_s$ is the number of particle degrees of freedom for a given spin $s$ with rest masses less than the black hole temperature $T_H$.
	
	The total power loss of a higher-dimensional Schwarzschild black hole  can be sufficiently approximated in the geometric optics limit.
	The dependence on the Hawking temperature $ T_H $ and the Schwarzschild radius $r_S $ is given by:
	
	\begin{equation}
		-\frac{dM}{dt} \approx c_n T_H^{n+4} r_S^{n+2}~,
	\end{equation}
	where $ c_n $ is a dimensionless constant that depends on the number of extra dimensions $ n $ and the spectrum of emitted particles. The above formula can be reformulated in terms of temperature as follows:
	\ba
	-\frac{dM}{dt}
	&=& c_n \left(\frac{n+1}{4 \pi}\right)^{n+2} T_H^2~.\label{IntdMdt}
	\ea
 Using the  formulae (\ref{TempBH}) and  (\ref{rsradius})
we express $T_H$ in terms of the BH mass  and then  integrate (\ref{IntdMdt}) to obtain
	\be 
	\tau\approx \frac{1}{c_n M_{*}}\left(\frac{4 \pi}{n+1}\right)^{n+4}  \frac{n+1}{n+3}
	\left(\frac{ 8\Gamma
		\left(\frac{n+3}{2}\right)}{(n+2) \pi ^{\frac{n+1}{2} }}\right)^{\frac{2}{n+1}}
	\left(\frac{M}{M_{*}}\right)^{\frac{n+3}{n+1}}
	\ee 
	
	In standard four-dimensional spacetime, i.e., when $n=0$, primordial black holes  are leading dark matter candidates, however, they are constrained  within the asteroid-mass window.  In particular  from observations of extra-galactic and $\gamma$-ray backgrounds,  the allowed range  is  $10^{15}$-$10^{21}$ g (see figure~\ref{PBHlimits} and ref~\cite{Villanueva-Domingo:2021spv}). Smaller PBHs with masses  $\lesssim 10^{15}$ g are excluded as they would have evaporated by now due to Hawking radiation.

	For  the case of one extra dimension, $n=1$, and $M_*\approx 10^{10}{\rm GeV}$  which is of our interest, we obtain 
	\be 
	\tau\approx	\frac{128 \pi^4 }{3 c_1}\left(\frac{ M_{BH}}{ M_{*}}\right)\frac{1}{ M_{*}}\approx \frac{0.0725}{c_1}\left(\frac{ M_{BH}}{2\times 10^{10}{\rm g}}\right) (13.8 \times 10^9) { \rm y}
	\ee 
	where $13.8 \times 10^9 { \rm y}$ is the age of the Universe. 
	As a result, taking $c_1\approx 0.04$, the range of the PBHs which survive until now is\cite{Anchordoqui:2022txe,Anchordoqui:2024akj}
	\be 
	10^{11}{\rm g} \lesssim M_{BH} \lesssim 10^{21} {\rm g} 
	\ee

	The above findings have important	implications for Hawking radiation and dark matter.  The analysis of Ref.~\cite{Anchordoqui:2024akj},  shows that for a given BH mass $ M_{\text{BH}} $, the horizon radius $ r_H$ in the higher-dimensional scenario is larger than its four-dimensional counterpart. This is because the gravitational force becomes stronger at short distances, allowing the BH to trap light within a wider area.
	This increased size directly leads to a lower Hawking temperature, since the temperature is inversely proportional to the horizon radius ($ T_H \propto r_H^{-1} $). Furthermore, the total power of Hawking radiation is proportional to $ T_{\text{BH}}^2 $. A lower temperature therefore results in a significantly diminished radiation rate.
	
	The combined effect is a substantial increase in the black hole's lifetime. A higher-dimensional BH evaporates much more slowly than a standard four-dimensional BH of the same mass. This implies that  lighter  BH masses  could constitute a viable dark matter  candidate. Primordial black holes  with masses that would have evaporated away in a four-dimensional universe could potentially survive until the present day if they inhabit a spacetime with compact extra dimensions, thereby opening a new window for PBHs to account for the dark matter.

	\section{Rotating  Black Holes in Higher Dimensions}

	Having summarized the properties of non-rotating black holes, we proceed to  a natural extension and study the physics  of a higher-dimensional black hole  endowed with spin.
	Such an object  is described by the Kerr solution~\cite{Kerr:1963ud} which has become the standard model for rotating  black holes. 
	
	A rotating black hole is characterized by its mass $M$, angular momentum $J\propto  M a$, and -in principle- its charge $Q$. However, for the present context we consider that the BH is rapidly losing its charge $Q$ via Schwinger-type pair production~\cite{Schwinger:1951nm} and so $Q$ can be neglected. We therefore restrict our analysis to the two-parameter Kerr family defined by $M$ and $J$.

	We consider black holes  assumed to form on a four-dimensional brane with a total spacetime dimensionality of $D=4+n$, where, as before, $n$ represents the number of compact extra spatial dimensions.  Consequently, their initial state possesses only one dominant angular momentum component aligned with the brane, even though a higher-dimensional black hole could, in principle, have up to $[(n+3)/2]$ independent rotation planes (where the brackets $[\cdots]$ denote the integer part). The sub-dominant angular momenta can be justifiably neglected as they remain largely unchanged during the evolution. Furthermore, we adopt the well-motivated assumption that Hawking radiation occurs predominantly on the brane, rather than into the bulk.
	Later in this work, we will specialize to the scenario of a single extra dimension ($n=1$), subject to the size constraint given in Eq. (\ref{sizebound}).

	Next, within this context, we investigate whether rotating black holes have a longer lifetime than their higher-dimensional Schwarzschild counterparts. To answer this, we first analyze the event horizon's properties, as the horizon radius directly determines the Hawking temperature and, consequently, the black hole's evaporation rate. We will follow the notations and conventions of the reference~\cite{Ida:2005ax}.
	
	\subsection{The Event Horizon of a Higher-Dimensional Rotating Black Hole}
	
	In four-dimensional Einstein's gravity, a rotating  BH has an event horizon located at $ r_H = M + \sqrt{M^2 - a^2}$ (in units where $ G = c = 1 $), which is smaller than the Schwarzschild radius $ r_S = 2M $, as a consequence of frame-dragging. As we are interested in the generalization of this geometry to $ D = 4 + n $ dimensions, the appropriate metric is that of a Myers-Perry  black hole~\cite{Myers:1986un}, which can be conveniently expressed in Boyer-Lindquist coordinates~\cite{Boyer:1966qh}. Assuming that PBHs  are restricted on a  3-brane  they have a non-zero impact parameter along the brane. Therefore, as  already  explained earlier,  there is only one non-zero angular parameter about an axis in the brane. 
	In the subsequent analysis, we fix the angular coordinates associated with the extra dimensions to the equatorial plane, setting $\theta_j = \pi/2 $ for $ j = 2, 3, \dots, n+1$, so the higher-dimensional Myers-Perry metric reduces to an effective four-dimensional line element given by:
	\begin{equation}
		\begin{split}
			ds^2 &= \left(1 - \frac{\mu}{\Sigma(r,\theta) \,r^{n-1}}\right)dt^2 + \frac{2 a\mu \sin^2\theta}{\Sigma(r,\theta) \,r^{n-1}} \, dt \, d\varphi - \frac{\Sigma(r,\theta)}{\Delta(r)} dr^2 - \Sigma(r,\theta) d\theta^2 \\
			&\quad - \left( r^2 + a^2 + \frac{a^2\mu \sin^2\theta}{\Sigma(r,\theta) r^{n-1}} \right) \sin^2\theta \, d\varphi^2,
		\end{split}
		\label{BraneMetric}
	\end{equation}
	where
	\begin{equation}
		\Delta(r) = r^2 + a^2 - \frac{\mu}{r^{n-1}}, \quad \Sigma(r,\theta) = r^2 + a^2 \cos^2\theta.
	\end{equation}
	This metric resembles the standard four-dimensional Kerr solution but incorporates the dependence on the number of extra dimensions $ n $, primarily through the modified gravitational potential $ \mu / r^{n-1}$.

	The condition \( \Delta(r) = 0 \), which determines	 the event horizons 	 yields:
	\begin{equation}
		r_H^{n+1} (1 + a_*^2) = \mu, \quad \text{where} \quad a_* \equiv \frac{a}{r_H}.
	\end{equation}
	Solving for the horizon radius \( r_H\), we find:
	\begin{equation}
		r_H = \left( \frac{\mu}{1 + a_*^2} \right)^{\frac{1}{n+1}}
		= \frac{r_S}{\left(1 + a_*^2\right)^{\frac{1}{n+1}}}~,
		\label{SchwaR}
	\end{equation}
	where in the second equality the parameter   $\mu$ has been replace by the Schwarzschild radius through their relation $ r_S = \mu^{1/(n+1)}$.  
	Remarkably, in the last formula of equation~(\ref{SchwaR}) the denominator is bigger than one which demonstrates that the angular momentum leads to a smaller horizon radius in higher dimensions.
	
	In addition, we should point out that in cases with $n=0$ and $n=1$, there is a maximum possible value of the parameter $a$, otherwise there are no solutions of $\Delta=0$ and there is no  horizon  to shield the singularity at $r=0$. For $n=1$ in particular  $\Delta=0$ implies
	\be 
	r_H=\sqrt{\mu-a^2}
	\label{5Dhorizon}
	\ee 
	 thus a horizon exists as long as $a<\sqrt{\mu}$.

	\subsection{Mass, Angular Momentum, and Temperature}
	
	The physical mass $ M_{\text{BH}} $ and angular momentum $ J $ of the black hole are derived from the asymptotic structure of the metric~(\ref{BraneMetric}). They are given by:
	\begin{align}
		M_{\text{BH}} &= \frac{(n+2)}{16\pi G_{4+n}} A_{n+2} \, \mu, \\
		J &= \frac{2}{n+2} M_{\text{BH}} a,
		\label{JKerr}
	\end{align}
	where $G_{4+n} $ is the gravitational constant in $ 4+n $ dimensions, and 
	\be
	 A_{n+2} =\frac{ 2\pi^{(n+3)/2}}{ \Gamma\left(\frac{n+3}{2}\right)}~
	 \label{AreaN}
	 \ee 
	 is the area of a unit $(n+2)$-sphere.
	
	The Hawking temperature, a crucial parameter for evaporation, is given by:
	\begin{equation}
		T_{\text{BH}} = \frac{1}{4\pi r_H} \frac{n+1+(n-1)a_*^2}{1+a_*^2}\label{TBHKerr} 
	\end{equation}
	Observe that the relationship between $r_H $, $ a_* $, and the temperature $T_{\text{BH}}$ is non-trivial.

	\subsection{ Field Equation and Propagation on the Brane}

	We now analyze the propagation of various fields within the gravitational background which it described by the metric~(\ref{BraneMetric}), induced on the brane.  
	Then, a solution of the wave equations can be obtained by decoupling the  resulting partial differential equation into radial and angular parts, by  employing the well known Ansatz for the separation of variables~\footnote{The separability of the equations for higher spins can be worked out in the context of Newman-Penrose formalism~\cite{NP}. See also~\cite{Ida:2005ax}. }
	\begin{equation}
		\phi(t, r, \theta, \varphi) = e^{-i\omega t} e^{im\varphi} R_{s\ell m}(r) \, {}_s S_{\ell m}(\theta, a\omega),
	\end{equation}
	Here $ \omega $  is the field's frequency, $ m $ is the azimuthal quantum number, and $ s$ denotes the spin weight. 
	Then, the dynamics of brane-localized fields with spin\footnote{At the classical level the Myers Perry  solution is valid for any spin. For gravitons emitted in the bulk, however, this is not true since emission of gravitons is a quantum effect and modifications are required.  In the semi-classical framework Hawking radiation is calculated in a fixed background and back-reaction is ignored.} $s = 0, 1/2, 1$ in the rotating black hole background are governed by the radial and angular equations which are:
	\begin{align}
		\frac{1}{\sin\theta}\frac{d}{d\theta}\left(\sin\theta\frac{dS}{d\theta}\right) + \left[(s-a\omega \cos\theta)^2 - (s\cot\theta + m\csc\theta)^2 - s(s-1) + A\right]S &= 0, \label{eq:angular} \\
		\Delta^{-s}\frac{d}{dr}\left(\Delta^{s+1}\frac{dR}{dr}\right) + \left[\frac{K^2}{\Delta} + 2 i s \omega r - a^2\omega^2 + 2 m a \omega - A\right]R &= 0, \label{eq:radial}
	\end{align}
	where $K = (r^2 + a^2)\omega - m a$, and $A$ is the angular eigenvalue. 
	The angular equation yields the eigenfunctions ${}_s S_{\ell m}(\theta, a\omega)$ -known as spin-weighted spheroidal harmonics -with eigenvalue $A = l(l+1) - s(s+1) + \mathcal{O}(a\omega)$. In the limit of slow rotation or low frequency ($ a\omega \ll 1 $), these harmonics reduce to the standard spin-weighted spherical harmonics ${}_s Y_{\ell m}(\theta, \varphi) $~\cite{Goldberg:1966uu}.
	
	The search for an exact solution to the radial equation is more challenging. A standard technique is to find analytic solutions in the  asymptotic regimes  near the event horizon (\( r \approx r_H\)) and at spatial infinity (\( r \gg r_H \)). These solutions are then matched in an intermediate region to construct a complete, approximate solution valid for all \( r \), which allows for the calculation of the greybody factor \( {}_s\Gamma_{\ell m} \),  as function of the parameters \( \omega, \ell, m, a, \) and \( n \)~\cite{Ida:2005ax}.\footnote{More precisely,  the greybody factor, ${}_s\Gamma_{lm}$, is the probability that a particle (Hawking radiation) emitted from near the horizon escapes to infinity without being backscattered by the black hole's spacetime curvature. It is equal to absorption probability   that an incoming wave (from infinity) of frequency $\omega$ and mode $(l, m)$ is absorbed by the black hole.}

	To proceed with the above described method of solution, the following dimensionless variables are introduced:
	\begin{align}
		\xi &= \frac{r - r_H}{r_H}, & \tilde{\omega} &= r_H \omega, & \tilde{Q} &= \frac{\omega - m\Omega}{2\pi T_H} = (1 + a_*^2)\tilde{\omega} - m a_*~, \label{xivariable}
	\end{align}
	where \( \Omega = a / (r_H^2 + a^2) \) is the angular velocity of the black hole, and \( T_H \) is the Hawking temperature, defined in Eq.~\eqref{TBHKerr}.
	Then, the radial equation becomes~\cite{Ida:2005ax}:
	\begin{equation}
		\xi^2(\xi+2)^2 R_{\xi\xi} + 2(s+1)\xi(\xi+1)(\xi+2) R_{\xi} + \tilde{V} R = 0~,\label{xiradial}
	\end{equation}
	where $\tilde{V}$ is the potential which, near the horizon limit
	(\(\tilde{\omega}\xi \ll 1\)),  takes the simplified form, 
	\[ \tilde{V}\approx \tilde{Q}^2-2is(\xi+1)\tilde{Q} -A_0\xi(\xi+2)+{\cal O}(\xi\tilde{\omega})~.\]
	Then, the solution satisfying the purely ingoing boundary condition is proportional to the ${}_2F_1$ hypergeometric function:
	\begin{align}
		R_{\mathrm{NH}} = \left(\frac{\xi}{2}\right)^{-s - i\tilde{Q}/2} \left(1 + \frac{\xi}{2}\right)^{-s + i\tilde{Q}/2} \times {}_2F_1\left(-l - s, l - s + 1, 1 - s - i\tilde{Q}; \frac{\xi}{2}\right). \label{NHsolution}
	\end{align}
	
	Similarly, the solution in the far-field  limit $(\xi \gg 1 + |\tilde{Q}|)$,  is written in terms of Kummer's confluent hypergeometric function ${}_1F_1$. The expression is:
	\ba
	R_{\mathrm{FF}} &=& B_1 e^{-i\tilde{\omega}\xi} \left(\frac{\xi}{2}\right)^{l-s} {}_1F_1(l-s+1, 2l+2; 2i\tilde{\omega}\xi)\nonumber\\
	&+& B_2 e^{-i\tilde{\omega}\xi} \left(\frac{\xi}{2}\right)^{-l-s-1} {}_1F_1(-l-s, -2l; 2i\tilde{\omega}\xi)~,\label{eq:FF_solution}
	\ea
	where the coefficients $B_1$ and $B_2$ are determined by asymptotically matching the near-horizon  and far-field  solutions in the intermediate region:
	\begin{align}
		B_1&= \frac{\Gamma(2l+1)\Gamma(1-s-i\tilde{Q})}{\Gamma(l-s+1)\Gamma(l+1-i\tilde{Q})}\\
		B_2&=\frac{\Gamma(-2l-1)\Gamma(1-s-i\tilde{Q})}{\Gamma(-l-s)\Gamma(-l-i\tilde{Q})}
	\end{align}

	Extending the far field solution to infinity (\(\xi \gg 1/\tilde{\omega}\)) yields asymptotic ingoing and outgoing waves, \(R_{\infty} \sim Y_{\text{in}} e^{-i\tilde{\omega}\xi} \xi^{-1} + Y_{\text{out}} e^{i\tilde{\omega}\xi} \xi^{-2s-1}\) where:
	\begin{align}
		Y_{in}&=\nonumber \frac{\Gamma(2l+1)\Gamma(2l+1) \Gamma(1-s-i\tilde{Q})}{\Gamma(l-s+1)\Gamma(l+s+1) \Gamma(l+1-i\tilde{Q})}(-4i\tilde{\omega})^{-l+s-1}\\&+\frac{\Gamma(-2l-1)\Gamma(-2l) \Gamma(1-s-i\tilde{Q})}{\Gamma(-l-s)\Gamma(-l+s) \Gamma(-l-i\tilde{Q})}(-4i\tilde{\omega})^{l+s}~,\\
		Y_{out}&=\frac{\Gamma(2l+1)\Gamma(2l+1) \Gamma(1-s-i\tilde{Q})}{[\Gamma(l-s+1)]^{2}\Gamma(l+1-i\tilde{Q})}(-4i\tilde{\omega})^{-l-s-1}\nonumber \\
		&+\frac{\Gamma(-2l-1)\Gamma(-2l) \Gamma(1-s-i\tilde{Q})}{[\Gamma(-l-s)]^{2}\Gamma(-l-i\tilde{Q})}(-4i\tilde{\omega})^{l-s}~.
	\end{align}
	Implementing the technics developed in~\cite{Page:1976df}  the absorption probability  $ {}_s\Gamma_{\ell m}$ (aka greybody factor) is finally given by:
	\begin{equation}
		{}_s\Gamma_{\ell m} = 1 - \left| \frac{Y_{\text{out}}}{Y_{\text{in}}} \right|^2 = 1 - \left| \frac{1 - C}{1 + C} \right|^2, \label{eq:gamma_final}
	\end{equation}
	where the factor $C$ is given by:
	\begin{equation}
		C = \frac{(4 i \tilde{\omega})^{2l+1}}{4} \left( \frac{(l+s)!(l-s)!}{(2l)!(2l+1)!} \right)^2 \left( -l - i \tilde{Q} \right)_{2l+1}~, \label{eq:C_factor}
	\end{equation}
	and $(z)_n$ denotes the Pochhammer symbol. This result is valid in the low-frequency regime $\tilde{\omega} \ll 1$ since higher order $\tilde{\omega}$ terms in (\ref{eq:gamma_final}) have been neglected. 
	Thus, the graybody factors in~(\ref{eq:gamma_final}) have been expanded  to the appropriate order where subleading terms are not included~\cite{Ida:2005ax} 
	\be 
	\begin{split}
		_0\Gamma_{00}&=4\tilde{\omega}^{2}-8\tilde{\omega}^{4}+O(\tilde{\omega}^{2})~,
		\\
		_0\Gamma_{1m}&=\frac{4\tilde{Q}\tilde{\omega}^{3}}{9}(1+\tilde{Q}^{2})+O(\tilde{\omega}^{6})~,
		\\
		_0\Gamma_{2m}&=\frac{16\tilde{Q}\tilde{\omega}^{5}}{2025}(1+\frac{5\tilde{Q}^{2}}{4}+\frac{\tilde{Q}^{4}}{4})+O(\tilde{\omega}^{10})	~,
		\\
		_{1/2}\Gamma_{1/2m}&=\tilde{\omega}^{2}(1+4\tilde{Q}^{2})-\frac{\tilde{\omega}^{4}}{36}(1+4\tilde{Q}^{2})^{2}+O(\tilde{\omega^{6}})~,
		\\
		_{1/2}\Gamma_{3/2m}&=\frac{\tilde{\omega}^{4}}{36}(1+\frac{40\tilde{Q}^{2}}{9}+\frac{16\tilde{Q}^{4}}{9})+O(\tilde{\omega}^{6})~,
		\\
		_{1}\Gamma_{1m}&=\frac{16\tilde{Q}\tilde{\omega}^{3}}{9}(1+\tilde{Q}^{2})+O(\tilde{\omega}^{6})	~,
		\\
		_1\Gamma_{2m}&=\frac{4\tilde{Q}\tilde{\omega}^{5}}{225}(1+\frac{5\tilde{Q}^{2}}{4}+\frac{\tilde{Q}^{4}}{4})+O(\tilde{\omega}^{10})~.
		\label{Gamas}
	\end{split}
	\ee 
	Notice, however, that the above approximations may result to smaller greybody factors.  Nevertheless, this precision of the estimates of the lifetime of the PBHs is adequate for our purposes.

\begin{figure}[h]
		\centering
		\includegraphics[width=0.375\textwidth]{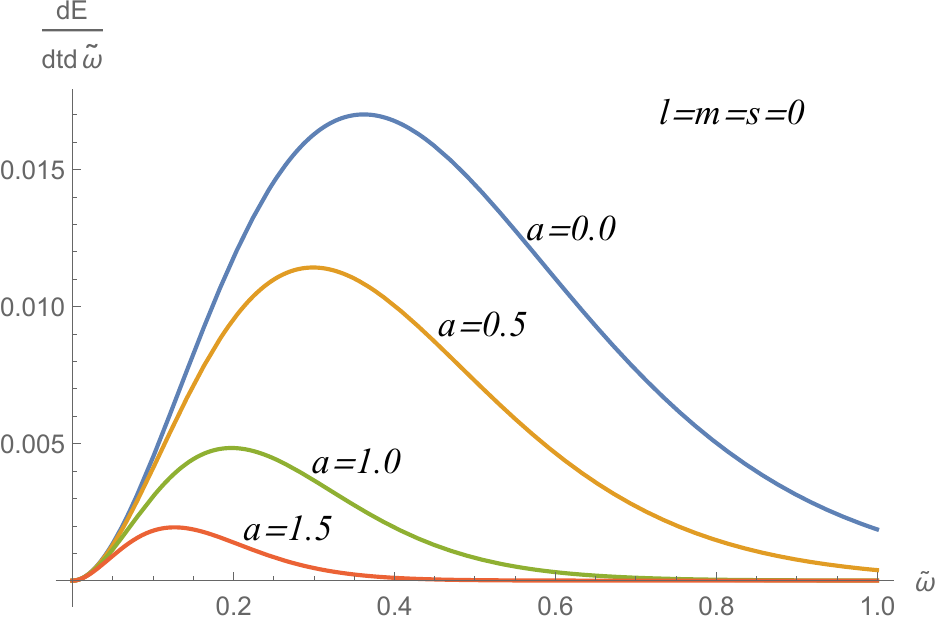} 
		\includegraphics[width=0.375\textwidth]{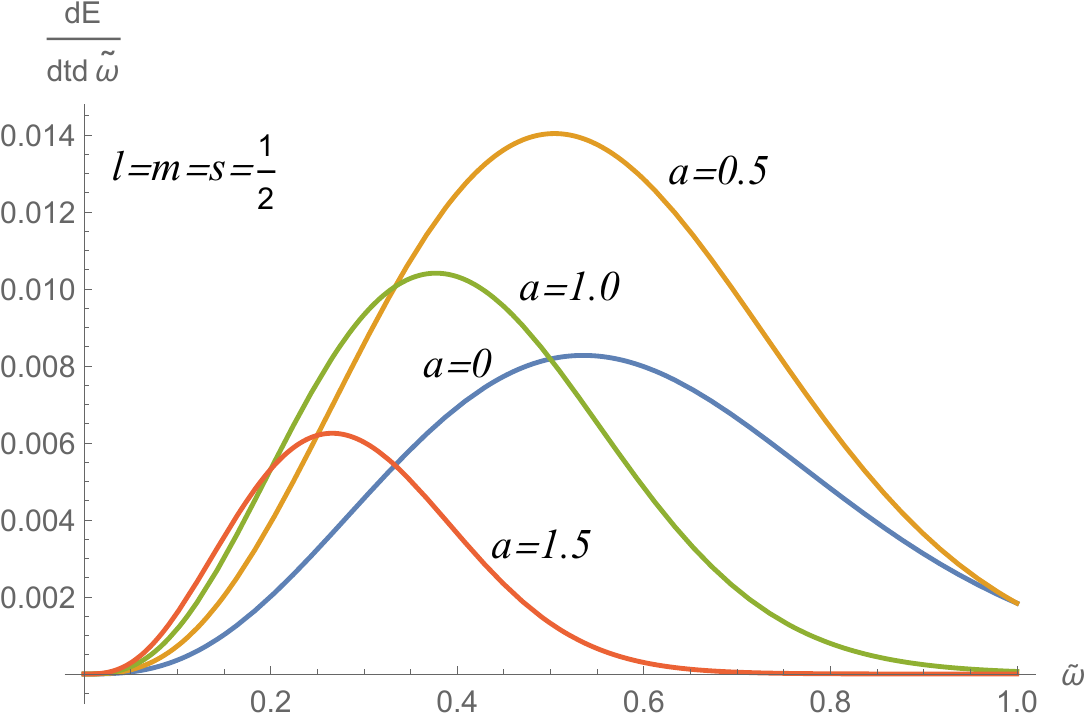} \\
		\includegraphics[width=0.375\textwidth]{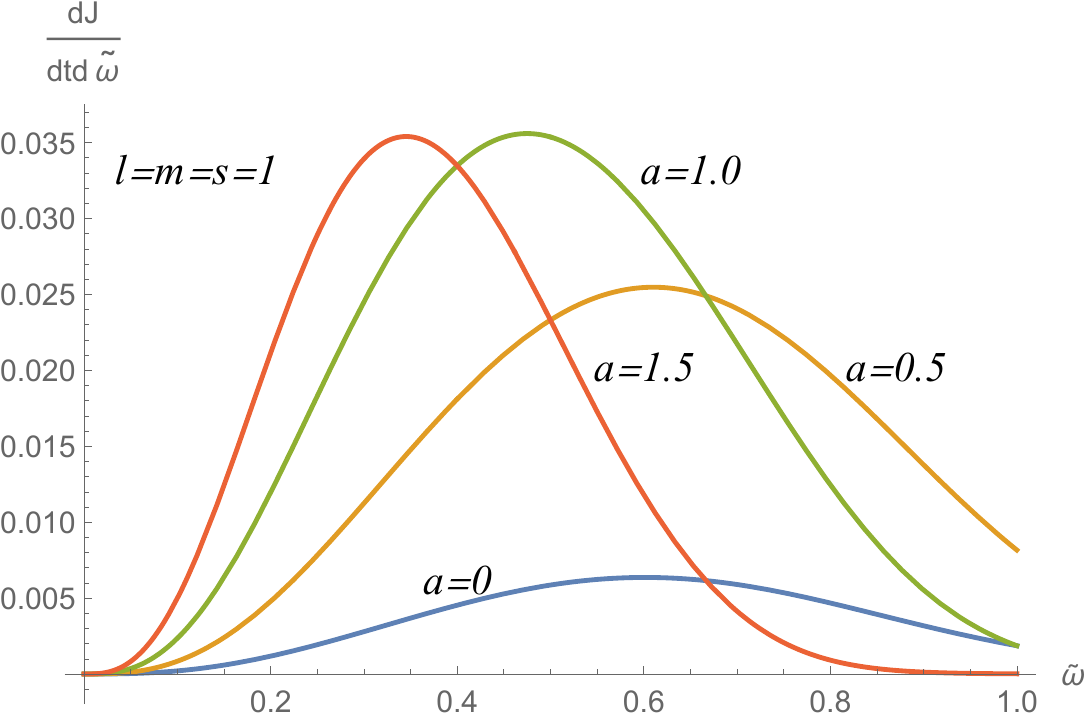} 
		\caption{Power Spectrum $\frac{dE_{slm}}{d\omega dt}$ vs $\tilde\omega$ for $l=m=s=\{0,1/2,1\}$. }
		\label{fig:plsm111}
	\end{figure}

	The Hawking radiation spectrum is given by the differential emission rate per unit frequency and unit time for a mode with quantum numbers $ (s, \ell, m)$:
	\begin{equation}
		\frac{d^3 E_{s\ell m}}{d\omega \, dt \, d\theta} = \frac{1}{2\pi} \frac{ {}_s\Gamma_{\ell m}(\omega) \, \omega}{\exp\left[ (\omega - m \Omega)/T_H \right] - (-1)^{2s}} \int_0^{2\pi} \left| {}_s S_{\ell m}(\theta, a\omega) \right|^2 d\varphi.
		\label{eq:emission_rate}
	\end{equation}
	The denominator $ \exp[ (\omega - m \Omega)/T_H ] - (-1)^{2s} $  accounts for the quantum statistics of the emitted particles, i.e., Bose-Einstein for bosons, and Fermi-Dirac for fermions.
	The integral over the azimuthal angle \( \varphi \) can be performed analytically for the spheroidal harmonics, yielding a function dependent only on $ \theta$. Explicit expressions for these $ \theta $-dependent factors for various modes can be found in the literature~\cite{Ida:2005ax, Press:1973zz}. The total emission rate is obtained by summing the contributions from all relevant particle species and quantum numbers $(s, \ell, m)$.
	In figure~\ref{fig:plsm111} we plot  the power spectrum $dE/(d\omega dt)$ for emitted particles with spins $s=0,1/2,1$.
	\paragraph{}
	The time dependence of the mass of the black hole is given by:
	\begin{equation}
		-\frac{dM}{dt} = \frac{1}{2\pi} \sum_{s, \ell, m} g_s \int_0^{\infty} \frac{ {}_s\Gamma_{\ell m}(\omega) \, \omega }{e^{(\omega - m\Omega)/T_H} - (-1)^{2s}} d\omega, \label{Μassrateloss}
	\end{equation}
	with $g_s$ being  the number of massless degrees of freedom at temperature $T_H$ with $g_0=4$, $g_{1/2}=90$ and $g_{1}=24$.

	Having calculated the expression for the greybody factor and the rate at which the black hole loses mass, in the next section we will calculate the lifetime of the 5D primordial rotating black hole.

	\section{5D Rotating Black Hole Evolution and Mass Loss}
	
	In this section we generalize previous results for the Schwarzchild BH by calculating the lifetime of a five-dimensional (\(n=1\)) rotating primordial black hole. We assume that Hawking radiation is dominated by emission of Standard Model particles confined to the brane, neglecting bulk fields like the graviton. The calculation proceeds by deriving an analytic expression for the greybody factors within a low-frequency expansion. Subsequently, the evolution of the coupled system of the mass and angular momentum of the rotating BH is investigated and the lifetime of PBHs are computed.

	\subsection{Angular momentum and   BH's lifetime } 
	
	For a rotating black hole, the angular momentum $J$ also evolves in time, and its rate of change is non-zero. This angular momentum loss rate couples to the mass loss rate given by Eq. (\ref{Μassrateloss}), 
	forming a system of coupled differential equations. A complete analysis must therefore account for the evolution of both $M$ and $J$. After performing the change of variables $\tilde{\omega} = \omega r_H$, $d\omega = d\tilde{\omega}/r_H$, and using the scaling property of the greybody factor, $\Gamma_{s,l,m}(\omega) = \Gamma_{s,l,m}(\tilde{\omega}, \tilde{a})$,   the two rates are given by:

	\begin{align}
		\frac{dM}{dt} &= -\frac{1}{2\pi r_H^2} \sum_{s,l,m} \int_0^\infty \frac{\tilde{\omega} \, \Gamma_{s,l,m}(\tilde{\omega})}{e^{(\tilde{\omega} - m\tilde{\Omega})/\tilde{T}} - (- 1)^{2s}} \, d\tilde{\omega}
		= -\frac{C_M({a_*})}{r_H^2}~, \label{eq:dMdt1}
		\\
		\frac{dJ}{dt}& = -\frac{1}{2\pi r_H} \sum_{s,l,m} \int_0^\infty \frac{m \, \Gamma_{s,l,m}(\tilde{\omega})}{e^{(\tilde{\omega} - m\tilde{\Omega})/\tilde{T}}- (- 1)^{2s}} \, d\tilde{\omega}
		= -\frac{C_J({a_*})}{r_H}~,\label{eq:dJdt1}
	\end{align}
	where $C_M$ and $C_J$ stand for the results with respect to $\tilde{\omega}$-integration and depend on $a_* = a/r_H$, and their numerical values will be given later.
	
	To a first approximation, we could integrate  the above relations, by initially making the assumption that the parameter $a_*$ remains approximately constant and equal to one. 
	A more precise estimate of the $M$ and $J$ evolutions, however, should take into account the simultaneous chance of both $M$ and $J$ quantities. We analyze this in the next sections.

	\subsection{The Angular momentum-BH mass relation}
	
	It is useful to directly extract the dependence of the angular momentum as a function of the black hole mass.
	The time variable can be eliminated by dividing the two equations, thereby deriving a relation between the black hole mass and angular momentum via an implicit function. We obtain
	\be 
	\frac{d J}{d M} =\frac{1}{\eta} r_H ~,\label{JfromM}
	\ee
	where the parameter $\eta$ is defined as follows:
	\be 
	\eta = \frac{C_M}{C_J} ~.\label{etadef} 
	\ee 
	Next, to further simplify the expressions, we introduce the parameters
	\be 
	\kappa =\dfrac{8 G_5 M}{3\pi},\; \lambda =\dfrac{9}{4 M^2}~,
	\label{KLdefns}
	\ee 
	so that the horizon radius (\ref{5Dhorizon}) is written as follows:
	\ba 
	r_H^2
	&=& \kappa \left(1 -\dfrac{\lambda}{\kappa}J^2\right)~.
	\ea 
	Notice that the condition $a<\sqrt{\mu}$,  derived from (\ref{5Dhorizon}) for the existence of a five dimensional black hole's  horizon,  here implies that 
	\be 
	J^2\le \frac{\kappa}{\lambda}\equiv \frac{32 G_5 }{27\pi}M^3\equiv J^2_{\rm max} ~,
	\label{Jmax} 
	\ee 
and	thus,  it defines a maximum value  $J_{\rm max}$ of the angular momentum. 
	
We proceed by substituting Eq.(\ref{KLdefns})  and $G_5=\frac{1}{8\pi M_*^3}$  to obtain 
	\be 
	r_H = \dfrac{1}{\pi M_*}\sqrt{\dfrac{M}{3M_*}}
	\sqrt{1-\mathcal{L}^2}~,\label{rHreparametrized} 
	\ee 
	where 
	\be 
	\mathcal{L}\equiv \mathcal{L}(M,J)= \left(\dfrac{M}{3M_*}\right)^{\frac 32}\dfrac{\pi }{2}J~.\label{Ldefinition} 
	\ee 
	Observe that, in this new parametrization the condition (\ref{Jmax})
	translates to ${\cal L}\le 1$.

	Plugging in the  definitions (\ref{rHreparametrized}) and  (\ref{Ldefinition}) into the equation (\ref{JfromM}) and separating variables we get
	\be 
	\frac{d\mathcal{L}}{\sqrt{1-\mathcal{L}^2}-\eta \mathcal{L}}
	=\frac{3}{2\eta} \frac{d M}{M}~.
	\ee 
Integration of the latter  by parts yields 
	\be 
	f(\mathcal{L})-f(\mathcal{L}_0)=\frac 32\left(\eta+\frac{1}{\eta}\right) \ln\frac{M}{M_0}~, \label{L_Mrelation} 
	\ee 
where the function $f(\mathcal{L})$ is given by
	\ba
	f(\mathcal{L})
	&=& \sin ^{-1}(\mathcal{L})-\eta  \log \left(\sqrt{1-\mathcal{L}^2}-\eta \mathcal{L}\right)~,\label{FL_sol}
	\ea 
	whilst  the constant of integration -denoted with $M_0$- stands for  the initial BH mass. 
	
	The logarithmic term in the evolution equations reveals that the dynamics primarily depends on the combination $\sqrt{1-\mathcal{L}^2} - \eta \mathcal{L}$, which has a clear physical interpretation as representing the balance between angular momentum loss and mass loss rates.
	
	We now return to the evolution equations for the BH mass and angular momentum. Using the $\mathcal{L}-M$ relation from equation (\ref{L_Mrelation}), we can decouple these equations. Focusing specifically on the angular momentum evolution, we begin with the definition from equation (\ref{Ldefinition}):
	\begin{equation}
		J = \frac{1}{\xi} M^{3/2} \mathcal{L}, \quad \text{with} \quad \xi = (3M_*)^{3/2} \frac{\pi}{2}.
	\end{equation}
	Differentiating this expression with respect to time and rearranging terms yields:
	\begin{equation}
		\label{Lrate}
		\begin{split}
			\frac{d\mathcal{L}}{dt} &= \frac{\xi}{M^{3/2}} \frac{dJ}{dt} - \frac{3}{2} \frac{\mathcal{L}}{M} \frac{dM}{dt} \\
			&= \frac{3}{2} \frac{C_M}{r_H^2} \frac{\mathcal{L}}{M} - \frac{\pi}{2} \frac{C_J}{r_H} \left( \frac{3M_*}{M} \right)^{3/2} \equiv \mathcal{F}(\mathcal{L}, M),
		\end{split}
	\end{equation}
	where in the last line we have defined the function $\mathcal{F}(\mathcal{L}, M)$ for later convenience.
	
	\noindent 
	The right-hand side of equation (\ref{Lrate}) depends only on $\mathcal{L}$ and $M$. However, using equations (\ref{L_Mrelation}) and (\ref{Ldefinition}), the black hole mass $M$ can be expressed exclusively as a function of $\mathcal{L}$:
	\ba
		\label{MofL}
		M\equiv M(\mathcal{L})&=& M_0 \left( \frac{\sqrt{1 - \mathcal{L}_0^2} - \eta \mathcal{L}_0}{\sqrt{1 - \mathcal{L}^2} - \eta \mathcal{L}} \right)^{\eta \zeta} e^{\zeta (\sin^{-1} \mathcal{L} - \sin^{-1} \mathcal{L}_0)} 
		\\
		&\equiv & M_0  \left( \frac{\sqrt{1 - \mathcal{L}_0^2} - \eta \mathcal{L}_0}{\sqrt{1 - \mathcal{L}^2} - \eta \mathcal{L}} \right)^{\eta \zeta}\left(\frac{{\cal L}+\sqrt{1+{\cal L}^2}}{{\cal L}_0+\sqrt{1+{\cal L}_0^2}}\right)^{\zeta},
	\ea
	where $\zeta = \frac{2}{3} \frac{\eta}{\eta^2 + 1}$.
	
	Substituting $M(\mathcal{L})$ into equation (\ref{Lrate}) allows us to express the evolution solely in terms of $\mathcal{L}$. The time $\tau_{\text{sd}}$ is then obtained by integrating:
	\begin{equation}
		\tau_{\text{sp}} = \int_{\mathcal{L}_0}^{0} \frac{d\mathcal{L}}{\mathcal{F}(\mathcal{L}, M(\mathcal{L}))}.
	\end{equation}

	The initial value of the angular momentum parameter, ${\mathcal{L}_0}$, is set by the initial conditions on $J$. Equation (\ref{FL_sol}) reveals that, for ${\mathcal{L}_0} < 1 / \sqrt{n^2 + 1}$, the  logarithmic term decreases monotonically with $\mathcal{L}$. Consequently, the duration of the spin-down phase -the time required for the black hole to eliminate its angular momentum- depends on the ratio $\eta = C_M / C_J$ and is therefore sensitive to the greybody factors. 	
	
	\begin{figure}[h]
		\centering
		\includegraphics[width=0.45\textwidth]{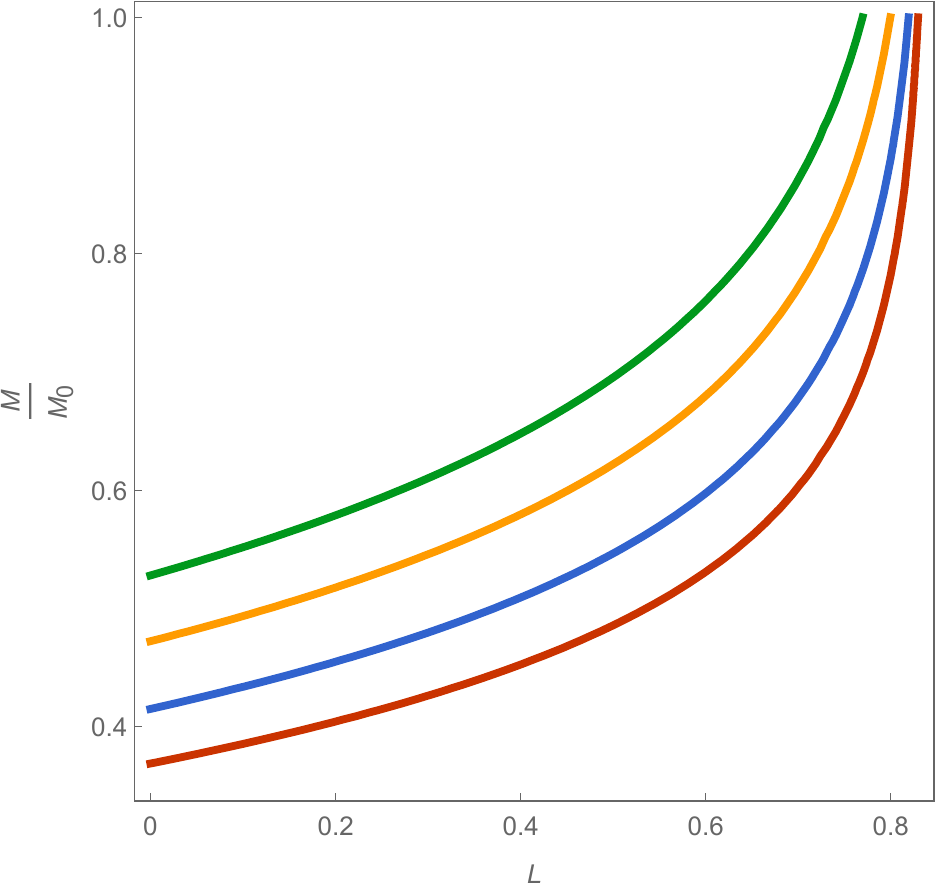}\\
			\includegraphics[width=0.4\textwidth]{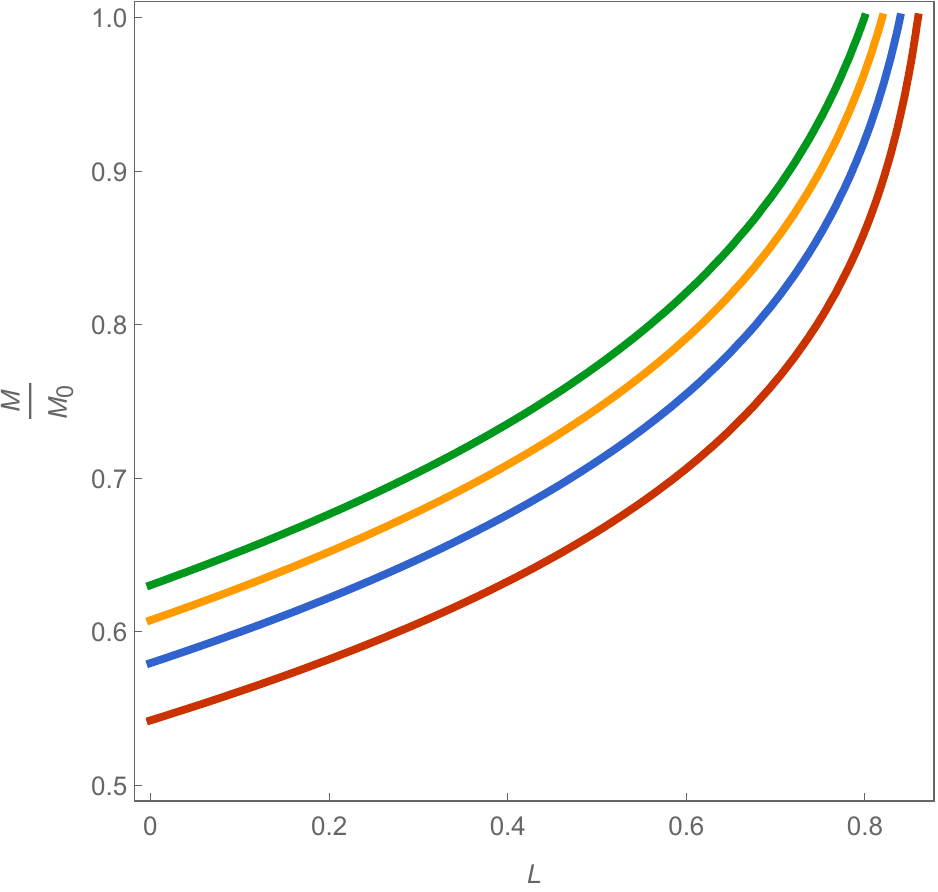}\;
	\includegraphics[width=0.4\textwidth]{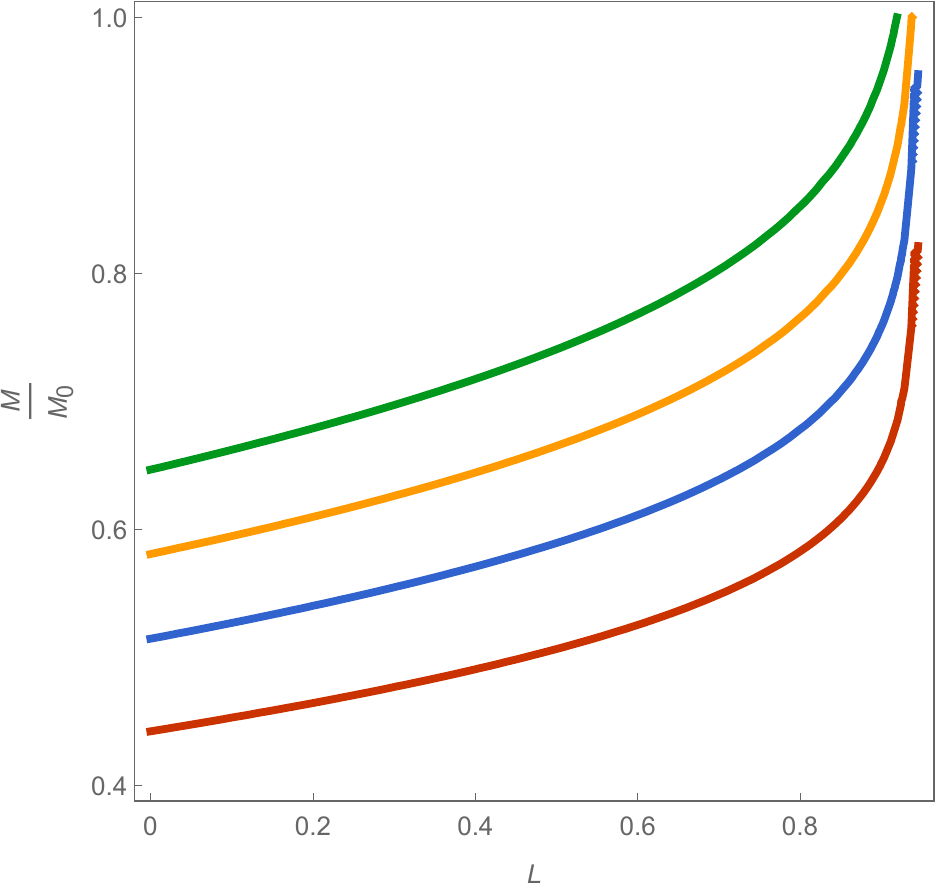}
		\caption{The ratio $M/M_0$  as a function of   $\mathcal{L}$ for the three cases of Table~\ref{xyz}. The upper plot corresponds to $\eta \approx 0.64, a_*=1/2$ and the curves are for four initial values in the range ${\cal L}_0=\{0.83-0.77\}$.  The  lower left plot is for   $\eta\approx 0.5, \alpha_*=1$ and  ${\cal L}_0=\{0.86-0.80\}$. Finally, the lower right plot is for $\eta\approx 0.35, a_*=3/2$ and ${\cal L}_0=\{0.944-0.92\}$.}
		\label{fig:LMcontours}
	\end{figure}
	 In figure~\ref{fig:LMcontours} we plot contours of the mass ratio $M/M_0$ of the rotating black hole as a function of its angular momentum ${\cal L}(J)$  for the three cases of $a_*$ and $\eta$ of Table~\ref{xyz}, and  several  initial conditions  of ${\cal L}$.  In each case, the maximum mass loss -at the time the angular momentum is eliminated- corresponds to the lower (red) curve and in all cases exceeds the $60\%$ of its initial mass $M_0$. In general, for values of $\mathcal{L}_0$ within 10\% of its maximum, the final mass loss is between $40\%$ and $60\%$ of the initial mass $M_0$.
	 
	 \noindent Upon examining the evolutionary tracks across all three panels of Figure~\ref{fig:LMcontours}, the following general trend emerges:
	 Black holes with higher initial angular momentum $\mathcal{L}_0$  exhibit a significantly longer Kerr phase, during which spin-down occurs gradually via  Hawking radiation. Consequently, a larger fraction of the initial mass is radiated away before the black hole transitions to the non-rotating Schwarzschild phase.

	\subsection{The lifetime of 5 dimensional rotating PBHs}
	
	The calculation of the lifetime for a five-dimensional rotating black hole is performed in two distinct phases. We first compute the spin-down time, which is the duration for the black hole to lose its angular momentum. We then calculate the lifetime of the resulting Schwarzschild black hole using the mass-loss relation (\ref{Μassrateloss}). Here, in this section, our approach relies on the key approximation that the mass decay rate is largely independent of the rotation parameter $a_*$.

	After performing the change of variables $\tilde{\omega} = \omega r_H$ (so that $d\omega = d\tilde{\omega}/r_H$) and utilizing the scaling property of the greybody factor, $\Gamma_{s,l,m}(\omega) = \Gamma_{s,l,m}(\tilde{\omega}, \tilde{a})$, the mass loss rate~\eqref{Μassrateloss} for each quantum field takes the generic form
	\begin{align}
		\frac{dM}{dt} & = -\frac{C_M(a_*)}{r_H^2}. \label{eq:dMdt1}
	\end{align}
	Here, we recall that  the coefficient $C_M(a_*)$ encapsulates the result of the integration over the greybody factors. The numerical values of $C_M,C_J$ for three specific spin parameters, $a_* = 1/2, 1, 3/2$, are presented in Table~\ref{xyz}.\footnote{Details of the numerical  calculations are included in the Appendix.}
	\begin{center}
		\begin{table}
			\centering
			\begin{tabular}{c|cccc}
				&$a_*$&$C_M$&$C_J$&$\eta$\\
				\hline 
			$1$  & $\frac{1}{2}$& 0.07&0.11&0.64 \\
			$2$  & $1$& 0.11&0.22&0.50 \\
			3&$\frac 32$&0.05&0.14&0.33\\
	\end{tabular}
\caption{\small  The coefficients $C_M,C_J$ and the ratio $\eta$ for three cases of $a_*$.  }
\label{xyz}
\end{table}
\end{center}

Considering now the specific case where $C_M \approx 0.11$, (see Appendix), the mass loss rate in Eq.~\eqref{eq:dMdt1} becomes:
\begin{align}
	\frac{dM}{dt} \approx  -\frac{0.11}{r_{H}^2}.
\end{align}
The horizon radius $r_H$ is related to the Schwarzschild radius $r_S$ in this scenario by:
\begin{equation}
	r_{H} = (1 + a_{*}^2)^{-\frac{1}{n+1}} r_{S},
\end{equation}
where from (\ref{rsradius}) the Schwarzschild radius $r_S$ for $n=1$ is:
\begin{equation}
	r_{S}^2 = \frac{1}{2\pi M_{*}}  \frac{8\Gamma(2)}{3}  \frac{M_{\text{BH}}}{M_{*}} =
	 \frac{4}{3\pi} \frac{M_{\text{BH}}}{M_{*}^2} ~.
	\label{eq:rs_def}
\end{equation}

Following Swampland arguments\cite{Palti:2019pca,vanBeest:2021lhn}, the fundamental Planck mass is set to $M_{*} \sim \SI{e10}{\giga\electronvolt}$. Substituting the expression for $r_S$ from Eq.~\eqref{eq:rs_def} into the mass loss rate and evaluating the constants yields a differential equation for the mass:
\begin{equation}
	\frac{dM}{dt} \simeq -5.68 \times 10^{29} \ \frac{\si{\giga\electronvolt^3}}{M}.
	\label{eq:dmdt_final}
\end{equation}

To compute the spin-down timescale $\tau_{\text{BH}}^{s.d}$, we integrate Eq.~\eqref{eq:dmdt_final}. Based on our previous assertion that the black hole loses approximately 60\% of its initial mass during this phase (see also~\cite{Ida:2006tf}), we find the spin-down lifetime scales as:
\begin{equation}
	\tau_{\text{BH}}^{s.d} \simeq 5.2 \times 10^{-15} \left( \frac{M_i}{\si{\gram}} \right)^2 \ {\rm y}.
	\label{eq:lifetime_result}
\end{equation}

	For an initial primordial rotating black hole mass of $M_i \simeq \SI{5e11}{\gram}$, the duration of the spin-down phase, calculated from Eq.~\eqref{eq:lifetime_result}, is:
	\begin{equation}
		\tau_{BH}^{s.d}\simeq 3.3\times10^9 { \rm y }.\label{time s.d}
	\end{equation}
Following this spin-down phase, the black hole loses its angular momentum and becomes non-rotating. Its subsequent evolution is then governed by the Schwarzschild-Tangherlini metric. The remaining mass $M_f$ at the beginning of this phase evolves according to the relation~\cite{Anchordoqui:2022txe}:
	\begin{equation}
		\tau_{BH}^{S.T}\simeq9\times10^{-15}\left(\frac{M_{f}}{\rm g}\right)^{2} \,{\rm  y}. 
	\end{equation}
	Assuming that the remaining black hole mass is $M_{f}\sim 1.5\times10^{11}$g then:
	\begin{equation}
		\tau_{BH}^{S.T}\simeq2\times10^8\, {\rm y}. \label{tims S.T} 
	\end{equation}
	Finally the lifetime of a primordial 5D rotating black hole with initial mass $M_{i}\sim5\times10^{11}$g is given by:
	\begin{align}
		\tau^{n=1}_{M.P}=\tau^{s.d}_{BH}+\tau_{BH}^{S.T}\simeq3.5\times10^{9}{\rm  y}  \label{lifetime}
	\end{align}

Our analysis reveals a suppressed Hawking radiation rate for five-dimensional black holes, thereby relaxing observational constraints from CMB, galactic bulge positrons, and isotropic photon backgrounds. Hence, this suppression opens the possibility for PBHs surviving today with mass as low as  $M_{BH}\gtrsim 10^{10}$ g to contribute to dark matter. Within this context, in~\cite{Anchordoqui:2022txe} in particular, it is examined whether in the 5D dark dimension scenario,  black holes extended  in the range  $10^{14}{\rm g} \lesssim M_{\rm BH}   \lesssim 10^{21}{\rm g}$ are sufficient to provide an all-dark-matter interpretation. 
Let us also point out that the calculated horizon radii align with the five-dimensional framework. For instance, a black hole with $M_{\text{BH}} \approx \SI{5e11}{\gram}$ has a radius of $r_H \approx \SI{5e-5}{\micro\meter}$, while one with $M_{\text{BH}} \approx \SI{e17}{\gram}$ yields $r_H \approx \SI{2e-2}{\micro\meter}$. 
Furthermore, the Hawking temperature of the lightest PBHs evaporating today, with masses $M_{\text{BH}} \sim \SI{e12}{\gram}$, is  the required  scale to potentially explain the Galactic \SI{511}{\kilo\electronvolt} gamma-ray line \cite{Anchordoqui:2022txe}.

	\section{The Memory Burden Effect and its Implications for Rotating Primordial Black Holes in Higher Dimensions} 
	
	In this section we estimate the lifetime of a rotating primordial black hole, incorporating the memory burden effect.
	As has been  emphasized in our previous analysis,  according to Hawking a black hole emits radiation with a purely thermal spectrum, dependent only on its mass and angular momentum. This radiation carries no information about the quantum state of the matter that formed the black hole, implying that information is permanently lost during evaporation, a direct violation of quantum unitarity, which dictates that information must be preserved.
	
	The ``memory burden effect'' offers a potential resolution to this paradox. According to this, as a black hole evaporates and loses mass, the information about its initial state is not lost but is instead imprinted on the remaining mass. This accumulating information acts as a stabilizing factor, progressively slowing the Hawking evaporation rate. Rather than evaporating completely, the black hole may transition into a long-lived remnant  from which information can be eventually released.
	
	This modified process of evaporation has profound implications for primordial black holes. By extending their lifetime, the memory burden effect could significantly increase the fraction of PBHs surviving to the present day, thereby strengthening their viability as a dark matter candidate. Recent work has explored this scenario in a standard 4-dimensional spacetime ~\cite{Dvali:2025ktz}. In this paper, we extend this analysis to the more complex and phenomenologically rich case of rotating primordial black holes in the presence of compact extra dimensions.
	
	We start by recapitulating the basics of the memory burden effect proposed in~\cite{Dvali:2018xpy}.
	Let $\hat{a}_k,\hat{a}^{\dagger}_k$ be the annihilation and creation eigenmode  operators satisfying the ordinary bosonic commutation relations 
	$[\hat{a}^{\dagger}_i,\hat{a}_j]=\delta_{ij}$ and 
	$[\hat{a},\hat{a}_j]=[\hat{a}^{\dagger}_i,\hat{a}^{\dagger}_j]=0$.
	From these we form the number operators $\hat{n}_k=\hat{a}^{\dagger}_k \hat{a}_k$   associated 
	with the eigennumber-sets $k$. Let also, the operator $\hnz$ represent the number of quanta in the central, 
	``master'' state of the system with energy  $\varepsilon_0$.
	Then, the memory burden effect can be captured by the Hamiltonian
	\be 
	\hat{\mathcal{H}}= \varepsilon_0 \hat{n}_0+\hat{\mathcal{E}}_K\sum_{k=1}^H \hat{n}_k~,
	\ee 
	where, $\hat{\mathcal{E}}_K$ is the energy gap of the $n_k$   given by~\cite{Dvali:2025ktz}
	\be \hat{\mathcal{E}}_K= (1-\hat{n}_0/N_c)^p\varepsilon_K~.\label{Egap}
	\ee 
	Observe that the energy gap $\hat{\mathcal{E}}_K$ is expressed in terms of the ratio $\hnz/N_c$ which sets the strength of attractive interaction between the master mode and the memory modes,  and the integer  $p$  which  for $p>1$  yields to a non-linear  interaction. 
	If the master state of the system is nearly empty, 
	$\hnz \ll N_c$, the factor $\left(1 - \hnz/N_c\right)^p \approx 1$, so the memory modes (\ref{Egap})  have their full energy cost $\hat{\mathcal{E}}_K\approx \varepsilon_K$ and  thus, storing information requires a lot of energy. On the contrary,
	when  $\hnz$ approaches $ N_c$, the factor $\left(1 - \hnz/N_c\right)^p$ tends to zero, and a local minimum of the energy is at $\hnz=N_c$  where the memory modes become effectively gapless.
	A possible  escape of  $n_{\ell}$ quanta, $\hnz \rightarrow N_c - n_{\ell}$, would imply a finite energy gap given by:
	\begin{equation}
		\hat{\mathcal{E}}_K = \left(1 - \frac{N_c-n_{\ell}}{N_c}\right)^p \varepsilon_K = \left( \frac{n_{\ell}}{N_c}\right)^p \varepsilon_K
	\end{equation}
	Thus, the system exhibits resistance against the escape of any quanta or particles, as the memory content $\sum \hnk$ reshapes the system's energy landscape to stabilize itself and any deviation from $n_0 = N_c$ creates an energy barrier, resisting further decay.
	This describes the ``burden'' of the information and constitutes the essence of the memory burden effect.
	
	In \cite{Dvali:2025ktz}, the memory burden effect is applied to black hole physics through a specific dictionary that maps the prototype model onto black hole parameters:
	\begin{equation}
		\varepsilon_0 = r_S^{-1}, \quad N_c = K = S, \quad N_m = S/2, \quad \varepsilon_K = S^{1/2} r_S^{-1},
	\end{equation}
	where $S$ is the Bekenstein-Hawking entropy and $r_S$ is the Schwarzschild radius. The exponent $p$ in the model, while treated as a free parameter, critically determines when the memory burden effect becomes significant.
	
	The onset of the memory burden phase is triggered after the black hole loses a critical fraction of its initial mass, given by~\cite{Dvali:2025ktz}
	\begin{equation}
		q \equiv \frac{\Delta M_{\text{crit}}}{M_0} \simeq \left(p \sqrt{S}\right)^{-\frac{1}{p-1}}.
	\end{equation}
	This relation reveals a key dependence on $p$:
	\begin{itemize}
		\item For $p=1$, we find $q \sim e^{-1}\sqrt{S}\sim \mathcal{O}(1)$, implying the memory burden effect dominates instantaneously, leaving no room for a semi-classical evaporation phase.
		\item For $p=2$, the critical mass loss is $q \sim S^{-1/2}$. This means the semi-classical description breaks down after a relatively short time $\tau \sim \sqrt{S} \, r_S$, after which the evaporation rate is strongly suppressed.
		\item For $p \gtrsim \ln S$, the critical mass loss is again $q \sim \mathcal{O}(1)$, delaying the onset of the memory burden phase until the black hole has lost approximately half of its mass.
	\end{itemize}
	Therefore, the value of $p$ dictates the evaporation timeline, controlling whether the transition from semi-classical to memory-burdened evaporation is early and sharp or late and gradual.

	\subsection{Memory burden for a 5D Kerr black hole}

	After the brief description of the memory burden effect, we are now ready to discuss its implementation in our  scenario.
	For the particular case of the rotating black hole, which is the focus of our work, we will assume that there is a first stage where the BH emits  Hawking (thermal) radiation. We will further suppose that this period  coincides with the phase of angular momentum loss, which is adequate for our purposes. Indeed, this is a reasonable assumption, as we have shown that the remaining mass at $J=0$ -depending on the initial condition- varies between 40\% and 60\% of the original. Consequently, in the second stage, the black hole becomes a Schwarzschild black hole, and the memory burden effect governs its subsequent evaporation.

	Following this reasoning, the Kerr phase of evaporation lasts for a period on the order of $10^9$ years, which is consistent with the timescales of the previous phases. Once the black hole loses its angular momentum and becomes a Schwarzschild black hole, the final stage of evaporation begins. Our calculation now focuses on this Schwarzschild phase, where we also incorporate the memory burden effect.

	\subsubsection{The  Schwarzchild phase }
	
	We already mentioned that, as the BH loses mass due to Hawking radiation, it gradually contracts while the accumulated information slows down the radiation, changing the expected decay rate. According to the memory burden scenario, at the end of this phase the BH transitions to a remnant or quantum critical phase where information is released. We will make the reasonable assumption that this happens around the time the rotation phase ends and therefore the remaining black hole has entered the Schwarzchild phase. We therefore focus on the mass evolution equation as a sufficient approximation for estimating the black hole lifetime.
	This is given by~\cite{Dvali:2018xpy}
	\be
	\frac{dM}{dt} = -\frac{1}{S^p}\frac{C_M }{r_H^2}~, \label{MBdMdt} \\
	\ee
	where, as before the coefficient $C_M$ stands for the $\tilde{\omega} $ integration and the entropy in the Schwarzchild case is given by $S = \frac{4\pi}3 M r_H$. Using the formula $r_H^2=\mu =\frac{8G_5}{3\pi}M$, we arrive at the equation
	\be 
	\frac{dM}{dt}= -\frac{C_S}{M^{\frac{3p}{2}+1}}
	\label{massloss}
	\ee 
	where 
	\be   
	C_S = C_M \left(\frac{4 \pi }{3}\right)^{-p}\left(\frac{8G_5}{3 \pi }\right)^{-\frac{p}{2}-1}\label{CSdef}
	\ee 	
	Separation of variables and integration yields 
	\ba 
	\tau^{(p)}&=&\left(\frac{4\pi}{3}\right)^p \left(\frac{8G_5}{3 \pi }\right)^{\frac{p}{2}+1}\frac{1}{C_M}\frac{2M^{2+\frac{3p}2}}{3p+4} \nn
	\\&=&\frac{2^{2 p+1}}{ 3^{\frac{3 p}{2}+1}}
	\frac{1
	}{ C_M\pi ^2(3 p+4)}	\left(\frac{M}{M_*}\right)^{\frac{3
		p}{2}+2}\frac{1}{ M_* }
\ea 
	where  in the last expression we substituted 
	$G_5=\frac{1}{8\pi M_*^3}~.$
	
	Table~\ref{MBcases} summarizes our computed primordial black hole  lifetimes for the parameter values $ p=0,1,2$. The standard scenario ($p=0$), with no memory burden, exhibits purely thermal Hawking evaporation; Setting $M_*\sim 10^{10} $ GeV and using the estimate  $C_M\approx 0.04$ (see~\cite{Anchordoqui:2022txe}), we obtain $\tau^{(p=0)}\approx 3\times 10^9 \left(\frac{M}{10^{12}g}\right)^2$y,  hence, PBHs of mass $\sim 10^{12}$g  have lifetimes exceeding the age of the universe and are viable dark matter candidates. For non-zero $p=1$  the memory burden regime onsets promptly, invalidating the semi-classical evaporation description. One finds  $\tau^{(p=1)}\approx 3.8\times 10^9\left(\frac{M}{12.74 g}\right)^{7/2}$.
	Consequently, for PBH masses at the epoch of our interest the evaporation rate is exponentially suppressed, leading to effective stabilization during the memory burden phase. The three cases are summarized in Table~\ref{MBcases}.

		
	\begin{center}
		\begin{table}
			\centering
			\begin{tabular}{|l|c|l|}
				\hline
				$p$&$\tau(M)$&Transition to M.B. \\
				\hline
				$0$  & $\frac{1}{6 \pi^2  }\frac{1}{C_M}\left(\frac{M}{M_*}\right)^2\frac{1}{M_*}\approx\; 2.78\times 10^{-15}\left(\frac{M}{{\rm g}}\right)^2 \rm y $
				&Hawking radiation\\
				$1$  & $\frac{8}{63\sqrt{3}\pi^2}\frac{1}{C_M}
				\left(\frac{M}{M_*}\right)^{7/2}\frac{1}{M_*}\approx \; 5.15\times10^{7}\left(\frac{M}{\rm g}\right)^{7/2}{\rm y} $ &Instantaneous MB effect\\
				2&$ \frac{1}{5\pi^2}\left(\frac{2}{3}\right)^4\frac{1}{C_M}\left(\frac{M}{M_*}\right)^{5}\frac{1}{M_*} \approx\; 1.17\times 10^{26}\left(\frac{M}{\rm g}\right)^5\rm y $&Semi-classical description  \\
				&& breaks down to MB phase\\
				\hline
			\end{tabular}
			\caption{\small  The lifetime of a rotating PBH as a function of its mass for three values of the exponent $p=0,1,2$ of the Entropy factor $1/S^p$ in the memory burden (MB) effect. $p=0$ corresponds to the standard Hawking radiation.  }
			\label{MBcases}
		\end{table}
	\end{center}	
		
Concluding this section, we wish to emphasize that the important consequence of the memory burden effect relevant to this work is that it predicts stable PBHs with masses well below $10^{10}$g that would otherwise have evaporated by today via Hawking radiation.
  In the classical scenario of Hawking evaporation, PBHs which are lighter that $M_{BH}\lesssim 5\times 10^{14}$g would have evaporated before the present epoch.  As a result, the evaporation particles (such as light SM states) can affect the early universe processes. In particular, observational constraints from light element abundances~\cite{Keith:2020jww}, such as deuterium and helium, put  constraints on evaporating BHs with mass range roughly   $\sim 10^8$g to $\sim 10^{13}$g. This is due to the fact that abundance of the emitted  high-energy particles would alter the expansion rate during the Big Bang Nucleosynthesis (BBN). 
	Therefore, if the memory burden mechanism is really effective, it  is especially significant since it predicts that Hawking evaporation for light PBHs  below $10^8$g comes to a standstill,  allowing them to survive until today and potentially contributing to the dark matter density without conflicting with BBN bounds.

	\section{Conclusions} 
	This work investigates whether higher-dimensional rotating primordial black holes  can account for the  dark matter of the universe in its entirety or only a fraction of it. The analysis is conducted within the framework of the dark dimension scenario, which considers a single extra compact dimension with a radius on the micrometer scale.
	
		We begin by establishing the essential properties of higher-dimensional black holes and reviewing the experimental bounds on their masses. We then compute the greybody factors for known rotating black hole solutions, which are crucial for accurately modeling Hawking evaporation. Subsequently, we analyze the coupled system governing the evolution of the black hole's angular momentum $J$, and mass $M $. We derive an analytical solution $ M(J) $, revealing that a black hole for initial $\mathcal{L}_0$ values close to maximum, loses more than 60\% of its mass during its spindown phase, a process lasting about $ 10^9 $ years. These results, however, exhibit a strong sensitivity to the greybody factors.
	Following spindown, the black hole enters a Schwarzschild phase and evaporates via Hawking radiation. We compute the lifetime of this second phase and find that, while it contributes to the total lifetime, it does not dramatically alter the lifespan established during the rotating phase. Crucially, within the dark dimension scenario, Hawking radiation is significantly suppressed. This allows PBHs with initial masses of the order $\gtrsim 10^{10} $ grams to survive until the present day, making them viable dark matter candidates.  
	
	Finally, we incorporate the memory burden effect, which further prolongs the black hole lifetime. Our comprehensive analysis demonstrates that higher-dimensional rotating PBHs, when considering both the suppressed evaporation and memory burden, remain compelling candidates to constitute the total dark matter content of the universe.

	\newpage 
	
	\section{
		APPENDIX 
	}

	\subsubsection*{The Mass Loss Rate for 5D Rotating Black Hole}
	
	The total rate of mass loss due to Hawking evaporation is found by summing over all particle species and quantum numbers:
	\begin{equation}
		-\frac{dM}{dt} = \frac{1}{2\pi} \sum_{s, \ell, m} g_s \int_0^{1} \frac{ {}_s\Gamma_{\ell m}(\omega) \, \omega }{e^{(\omega - m\Omega)/T_H} - (-1)^{2s}} d\omega, \label{massrateloss}
	\end{equation}
	where \(g_s\) counts the degrees of freedom for species of spin \(s\). Using the dimensionless variable \(\tilde{\omega} = r_H \omega\), this becomes:
	\begin{equation}
		\begin{split}
			\frac{dM}{dt}=& \nonumber-\frac{1}{2\pi}\int_{0}^{1}d\omega\sum_{s,l,m}\frac{_s\Gamma_{lm}\omega}{e^{2\pi[2\tilde{\omega}-m]}-(-1)^{s}}\\=&-\frac{1}{2\pi r_{h}^{2}}\left[g_{0}\int_{0}^{1}d\tilde{\omega}\sum_{m}\frac{_0\Gamma_{00}\tilde{\omega}}{e^{2\pi(2\tilde{\omega})}-1}\right.
			\\& \nonumber\left.  +g_{0}\int_{0}^{1}d\tilde{\omega}\sum_{m}\frac{_0\Gamma_{1m}\tilde{\omega}}{e^{2\pi(2\tilde{\omega}-m)}-1}+g_{0}\int_{0}^{1}d\tilde{\omega}\sum_{m}\frac{_0\Gamma_{2m}\tilde{\omega}}{e^{2\pi(2\tilde{\omega-m})}-1}\right. \\& \nonumber\left. +g_{1/2}\int_{0}^{1}d\tilde{\omega}\sum_{m}\frac{_{1/2}\Gamma_{1/2m}\tilde{\omega}}{e^{2\pi(2\tilde{\omega}-m)}+1}+g_{1/2}\int_{0}^{1}d\tilde{\omega}\sum_{m}\frac{_{1/2}\Gamma_{3/2m}\tilde{\omega}}{e^{2\pi(2\tilde{\omega}-m)}+1}+\right. 
			\\&\left.  g_{1}\int_{0}^{1}d\tilde{\omega}\sum_{m}\frac{_{1}\Gamma_{1m}\tilde{\omega}}{e^{2\pi(2\tilde{\omega}-m)}-1}+g_{1}\int_{0}^{1}d\tilde{\omega}\sum_{m}\frac{_{1}\Gamma_{2m}\tilde{\omega}}{e^{2\pi(2\tilde{\omega}-m)}-1}\right]  
		\end{split}
	\end{equation} 
	After numerical integration, the total emission rate can be expressed in the form  $$ -dM/dt = C_M / r_H^2~,$$ where $C_M$  is a dimensionless constant aggregating the contributions from all modes: \[C_M = \sum_{s, \ell, m} a_{s\ell m} g_s~.\]
	
	Calculating of the above integrals  we find that:
	\begin{align}
		\frac{dM}{dt}
		&\approx \frac{4(0.0008+0.0003+0.0005)+90(0.003+0.0017)+24(0.011+0.001)}{2\pi r_{h}^{2}}~, \nonumber\\
	\end{align}
	thus,
	\[ 		\boxed{\frac{dM}{dt} \,\approx\,-\frac{0.11}{r_{h}^2}}\]

\end{document}